\documentclass{ws-jcsc}

\usepackage[dvipsnames]{xcolor}
\usepackage{hyperref}
\usepackage{cleveref,multirow}
\usepackage{subfloat}
\usepackage{adjustbox}

\begin{document}

%
\catchline{}{}{}{}{}
%
\title{Area, Delay, and Energy-Efficient full Dadda Multiplier}

\author{Muteen Munawar$^1$, Zain Shabbir$^2$, Muhammad Akram$^3$}
\address{Department of Electrical, Electronics and Telecommunication Engineering,\\ University of Engineering and Technology,\\
Lahore, Punjab,
Pakistan\footnote{Muteen Munawar, Zain Shabbir and Muhammad Akram are with the Department of Electrical, Electronics and Telecommunication Engineering, University of Engineering and Technology, Lahore 54890, Pakistan}\\
$^1$muteen@seoultech.ac.kr\\ $^2$zain.shabbir@uet.edu.pk\\ $^3$muhammad.akram@uet.edu.pk}

\maketitle

\begin{history}
\received{(Day Month Year)}
\revised{(Day Month Year)}
\accepted{(Day Month Year)}
\end{history}

\begin{abstract}
The Dadda algorithm is a parallel structured multiplier, which is quite faster as compared to array multipliers, i.e., Booth, Braun, Baugh-Wooley, etc. However, it consumes more power and needs a larger number of gates for hardware implementation. In this paper, a modified-Dadda algorithm-based multiplier is designed using a proposed half-adder-based carry-select adder with a binary to excess-1 converter and an improved ripple-carry adder (RCA). The proposed design is simulated in different technologies, i.e., Taiwan Semiconductor Manufacturing Company (TSMC) 50 nm, 90 nm, and 120 nm, and on different GHz frequencies, i.e., 0.5, 1, 2, and 3.33 GHz. Specifically, the 4-bit circuit of the proposed design in TSMC's 50 nm technology consumes 25 uW of power at 3.33 GHz with 76 ps of delay. The simulation results reveal that the design is faster, more power-energy efficient, and requires a smaller number of transistors for implementation as compared to some closely related works. The proposed design can be a promising candidate for low-power and low-cost digital controllers. In the end, the design has been compared with recent relevant works in the literature.

\end{abstract}



\section{Introduction}
\label{Int}

A digital multiplier is an important building block of any logical processor in a digital system, and some of the main specifications of such systems like processing speed, power consumption, and energy efficiency highly depend on it \cite{ref1,ref2,ref3,ref4}. There is always a need to improve the performance of a multiplier to meet the requirements of fast and energy-efficient processes  \cite{ref5,ref6}. In digital image processing systems, convolution neural networks \cite{ref7,ref8}, and general-purpose processes, the performance of a multiplier must be considered, especially where mathematical data evaluation is of higher priority \cite{ref9,ref10}.

A general multiplication algorithm can be divided into three segments \cite{ref3,ref11,ref12}. The first part is where two \(n\)-bit numbers are given to the inputs of AND gates to generate the partial products (PPs). Then, these PPs layers are compressed using full adders (FAs) and half adders (HAs), unless only two layers of binary numbers remain. Finally, these two layers are added, usually by a RCA, in order to generate the final result of multiplication \cite{ref11,ref12}. The majority of the work in the literature is focused on the second segment, i.e., compressors, to reduce the overall delay, power consumption, and area of the multiplier \cite{ref11,ref12,ref13,ref14,ref15,ref42,ref45}.

In the literature, many publications can be found on the implementation of multiplier algorithms to reduce delay, power and energy consumption, and layout area. For example, in \cite{ref16} authors designed a 4-bit Dadda multiplier circuit using a reduced-split precharge-data driven dynamic sum logic (rspD3Lsum), which uses a lower number of transistors as compared to the traditional one. As a result, the power consumption and area of the chip were improved. In \cite{ref17}, the authors modified the carry-select adder (CSA) using a binary to excess-1 (BEC1) converter and used this circuit as a compressor to implement the circuit of the Dadda multiplier, which improved the speed, area, and energy as compared to the traditional CSA-based design. To reduce power and area, an optimized adder with pass transistor logic is used to design a dadda multiplier in \cite{ref18}, but the output voltage levels are not as strong as in CMOS logic. The authors proposed a Dadda circuit based on the carry look-ahead adder and optimized full adder in \cite{ref19}, which was implemented using complex cells in CMOS 65 nm technology. 

To increase the performance of the digital multipliers by improving the second stage of the multiplication, a number of works can be found in the literature. The parallel prefix \cite{ref20}, the approximate adder \cite{ref21}, attack-based, and the novel compressor \cite{ref22,ref23} are examples. These methods try to focus on the second step of multiplication, where different layers of PPs are reduced using compressors \cite{ref12}. Another type of multiplier implementation technique exists in the literature and is known as "approximate multipliers" \cite{ref24}. These multipliers are typically more area- and power-efficient than exact multipliers; however, they may contain errors and are best suited for error-tolerant applications. Various designs for these approximate multipliers have been proposed in the literature \cite{ref26,ref27,ref28,ref29,ref30,ref31,ref32,ref33}.

In this paper, a digital multiplier design is proposed that is based on the modified Dadda algorithm, also known as the "full Dadda algorithm" \cite{ref34}. To compress the layers of PPs in our design, a novel 3:2 adder has been proposed that is faster, more area-efficient, and more power-energy efficient as compared to other traditional FAs, i.e., carry look-ahead adders, simple FAs, CSAs, etc. In the end, an improved RCA \cite{ref35} has been used to calculate the final result. The design of the proposed circuits is validated in DSCH software, whereas the layouts are designed and simulated in Microwind software using Taiwan Semiconductor Manufacturing Company (TSMC) 50 nm, 90 nm, and 120 nm technologies. The simulation results show that the proposed design is better compared to the related works in terms of transistor count, energy efficiency, and delay.

This paper is organized as follows: the traditional Dadda algorithm and the full Dadda algorithm are explained in Section 2. In Section 3, the proposed design is explained in detail. Section 4 shows the simulation results, whereas Section 5 gives a brief conclusion to this paper.

\section{Full Dadda multiplier}
\label{FullD}

A Dadda multiplier is similar to a Wallace multiplier \cite{ref41}, but it is faster and needs a smaller number of gates for a multiplication operation. In Figures \ref{fig1a}, \ref{fig1b}, and \ref{fig1c}, the structure of a 4-bit Dadda multiplier can be seen. After generating 16 PPs of input bits, as shown in Figure \ref{fig1a}, PPs are arranged in a specific order, which can be seen in Figure \ref{fig1b}. \textcolor{black}{After arranging in this pattern, the Dadda multiplier uses the HAs and FAs to reduce the number of layers of PPs in such a way that each successive step reduces the number of layers by a factor of \(\frac{2}{3}\).} The Dadda multiplier tries to reduce the number of gates and input/output delay in each layer, whereas the Wallace multiplier attempts to reduce the layers as much as possible. More information on the traditional dadda multiplier can be found in the references \cite{ref36,ref37,ref38,ref39,ref40,ref41}.
\begin{subfigures}

\begin{figure}[h!]
\centerline{
\includegraphics[width=0.4\textwidth]{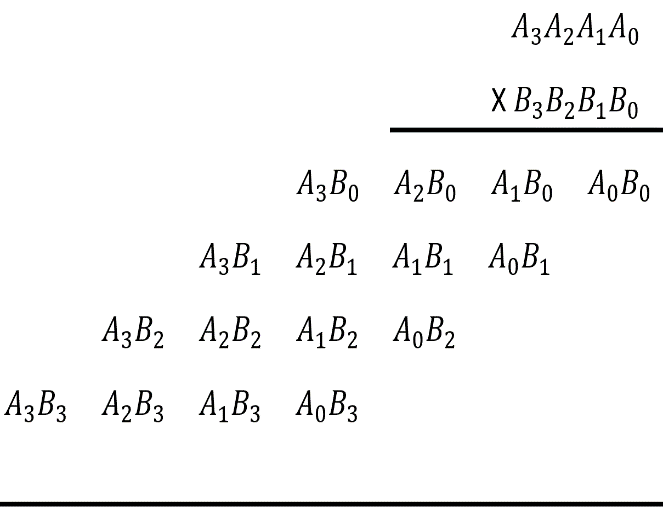}}
\caption{4x4 partial products generation [17].}
\label{fig1a}
\end{figure}


\begin{figure}[h!]
\centerline{
\includegraphics[width=0.4\textwidth]{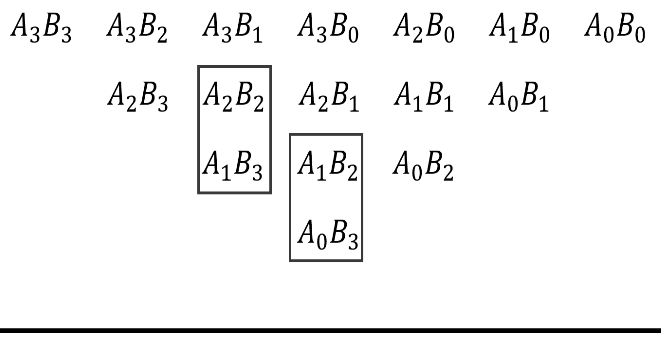}}
\caption{Dadda tree height [17].}
\label{fig1b}
\end{figure}

\begin{figure}[h!]
\centerline{
\includegraphics[width=0.4\textwidth]{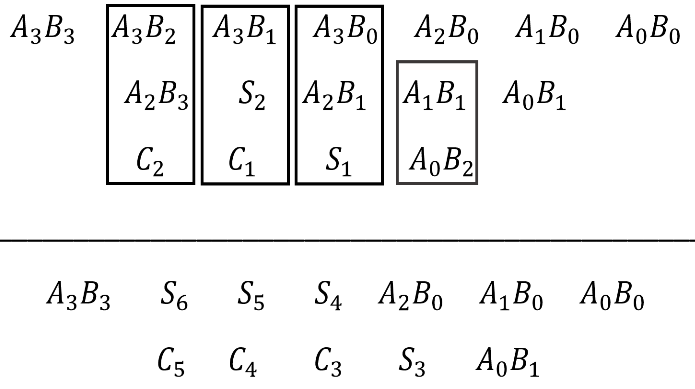}}
\caption{Final reduction stage [17].}
\label{fig1c}
\end{figure}

\end{subfigures}

The difference between a traditional Dadda and a full Dadda multiplier is that a full Dadda prefers to use FAs in the early stages of reduction while a simple Dadda uses HAs, except in the last stage where both are the same. The full Dadda provides a simple and more regular scheme as compared to the traditional one and uses a lower number of interconnections in circuit layout. As a result, full Dadda implementation requires less area as compared to a simple Dadda scheme. Some general equations to calculate the number of HAs, FAs, and carry-propagation adders (CPA) for an \(N\)-bit full Dadda multiplier are given below \cite{ref34}:

\begin{equation}
\label{eq1}
{\rm{HAs}} = N - 1,
\end{equation}

\begin{equation}
\label{eq2}
{\rm{FAs}} = (N - 1) \times (N - 3),{\rm{valid\,for\, }}N > 2,
\end{equation}

\begin{equation}
\label{eq3}
{\rm{Size\,of\,CPA}} = 2 \times (N - 2).
\end{equation}

More detail on the general rules and equations of the full Dadda multiplier can be found in \cite{ref34}. A comparison of the Dadda and full Dadda pattern, from \cite{ref34}, has been shown in Figure \ref{fig2a} and \ref{fig2b}, respectively.

\begin{subfigures}

\begin{figure}[h!]
\centerline{
\includegraphics[width=0.6\textwidth]{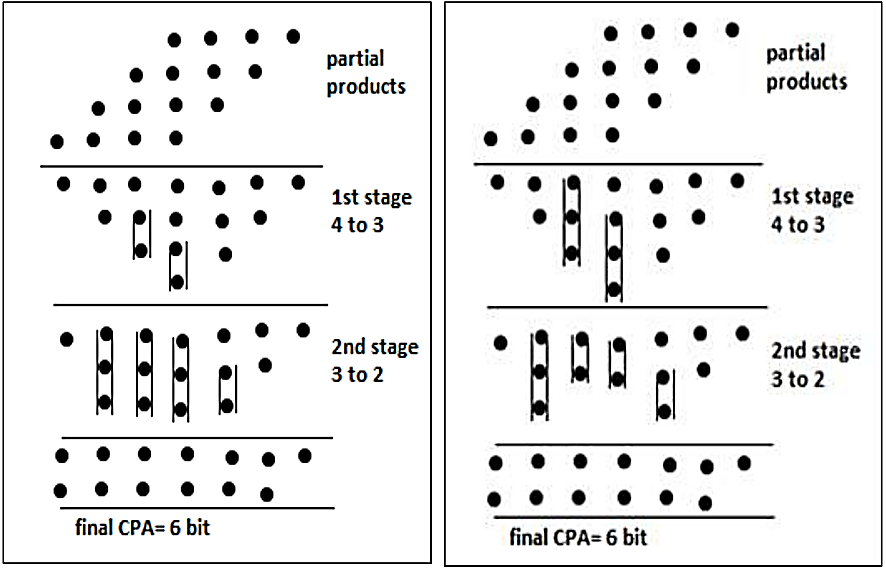}}
\caption{4-bit Dadda and full Dadda (from [34]).}
\label{fig2a}
\end{figure}

\begin{figure}[h!]
\centerline{
\includegraphics[width=0.6\textwidth]{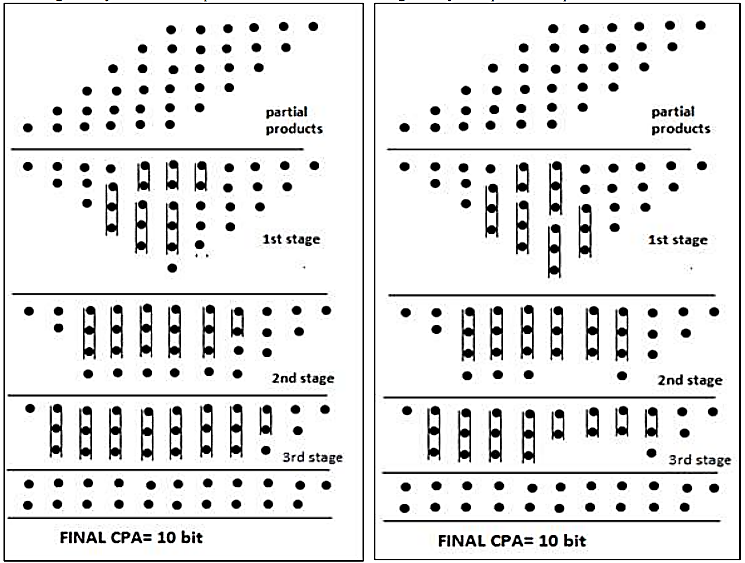}}
\caption{6-bit Dadda and full Dadda (from [34]).}
\label{fig2b}
\end{figure}

\end{subfigures}

\section{Proposed work}
\label{Proposed}

The implementation design of a full Dadda multiplier is proposed, where the reduction phase has been processed using our proposed adder. Before discussing the proposed adder, it is necessary to look at a traditional CSA (Figure \ref{fig3a}). A CSA calculates two results in parallel. One result considers that the incoming input carry bit is 0, while the second result considers carry bit 1. In this way, when the actual carry bit comes, this adder selects the corresponding result using a multiplexer (MUX) circuit. The benefit of using this technique is that the final result is much faster as compared to a simple carry propagation full adder because it does not have to wait for the incoming carry bit to start the addition process. However, the problem with this addition is that its implementation requires a large number of transistors, which results in a large area and high power consumption. In \cite{ref17}, the authors modified CSA with a BEC1 converter (CSA-BEC1) (Figure \ref{fig3b}) and used this adder in the implementation of the Dadda multiplier circuit. As a result, the overall circuit used a small area in the VLSI layout and consumed less power. As it can be seen in Figure \ref{fig3a}, the purpose of the FAs in the second row is just to give an addition result while considering that the incoming carry bit is 1. This series of FAs can be replaced by a BEC1 converter. A BEC1 adds one bit to the result while using fewer transistors. So instead of using a simple FAs series, the authors in \cite{ref17} used a BEC1 converter, and as a result, their multiplier achieved better results in terms of power, speed, and energy.

\textcolor{black}{This section is divided into two parts. In the first part, the design and workings of the proposed adder are explained. Consequently, based on the proposed adder, the design of the full Dadda multiplier is explained in the second part.} 

\begin{subfigures}

\begin{figure}[h!]
\centerline{
\includegraphics[width=0.5\textwidth]{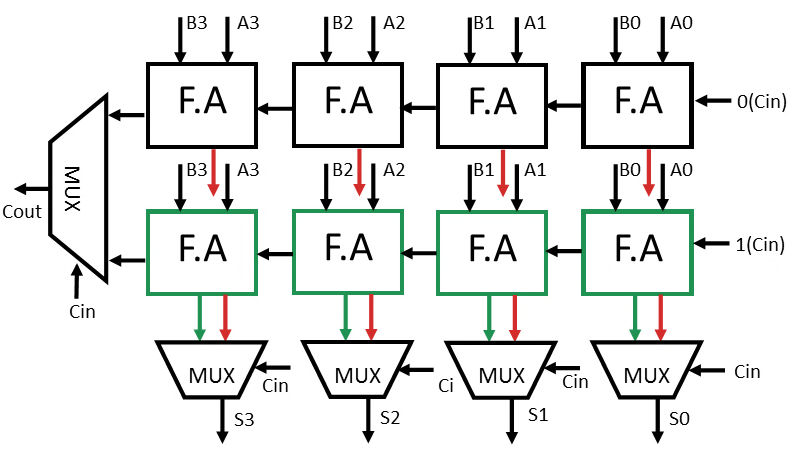}}
\caption{Carry select adder.}
\label{fig3a}
\end{figure}

\begin{figure}[h!]
\centerline{
\includegraphics[width=0.5\textwidth]{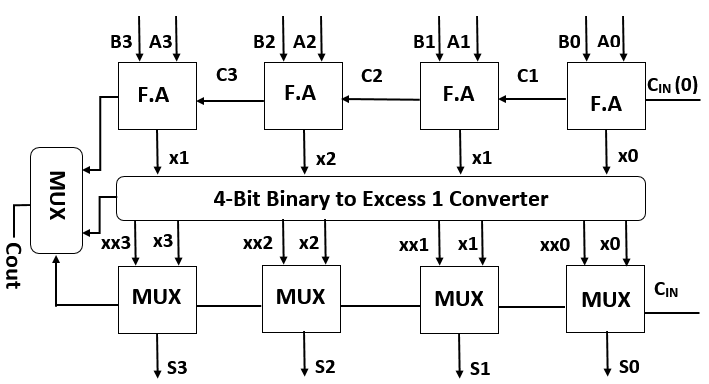}}
\caption{Carry select adder with binary to excess 1 converter (4-bit) [17].}
\label{fig3b}
\end{figure}

\end{subfigures}

\subsection{Proposed half adder based CSA-BEC1}

The CSA-BEC1 circuit is further modified in this paper, and it is then used in full daddy multiplier implementation to achieve better performance. A block diagram of the proposed modification is shown in Figure \ref{fig4a}. It can be seen that there is no FA in the proposed modification. Although all FAs have been replaced by HAs, this circuit still makes use of the speed advantages of a CSA. \textcolor{black}{Comparing this circuit with Figures \ref{fig3a} and \ref{fig3b}, it can be observed that there are no horizontal interconnections between the HA adders and the BEC1 circuit as there were in CSA and CSA-BEC1. Carry propagation overhead between FAs is shifted to MUX stages. Due to this overhead transfer, although the FAs can be replaced by HAs, the number of MUX is doubled; however, the overall number of transistors is reduced as compared to the simple CSA and CSA-BEC1}. Furthermore, the MUX circuit is implemented using pass transistor logic (PTL), which only needs 2 transistors for implementation (Figure \ref{fig4b}).

In summary, the FAs are replaced with HAs, which save at least 18 transistors per 1-bit in CMOS logic at a cost of only two transistors more in MUX. Hence, the total number of transistors is reduced, which will reduce the layout area and improve the delay and energy performance.

The workings of the improved adder, named HA-based CSA with BEC1 (HA-CSA-BEC1), are as follows: The HAs calculate the output considering that there is no input carry, i.e., 0, and similarly, the BEC-1 blocks calculate the addition results considering that the input is 1. When the actual carry comes to the input of MUX, the final selection of output is made based on the actual carry bit. Carry propagation takes place in the MUX stage rather than the FAs as it did in CSA and CSA-BEC1. However, this propagation of carry through MUXs is very fast as compared to CSA and CSA-BEC1 because there is only one transistor delay per 1-bit in the path of propagation.

\begin{subfigures}

\begin{figure}[h!]
\centerline{\includegraphics[width=0.5\textwidth]{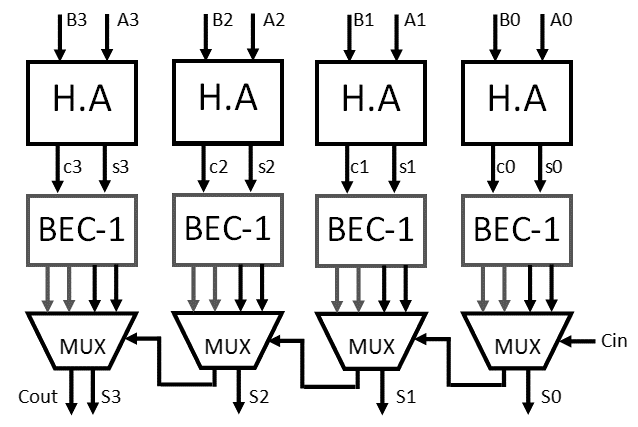}}
\caption{Proposed HA-CSA-BEC1 (4-bit).}
\label{fig4a}
\end{figure}

\begin{figure}[h!]
\centerline{
\includegraphics[width=0.18\textwidth]{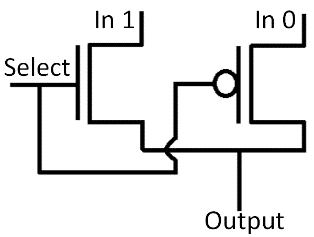}}
\caption{2:1 MUX using pass transistor logic.}
\label{fig4b}
\end{figure}

\end{subfigures}

A comparison of the total number of transistors is shown in Table \ref{table1}. The implementation of a 4-bit simple CSA, CSA-BEC1, and proposed HA-CSA-BEC1 with CMOS logic takes 256, 168, and 128 transistors, respectively. Details are given in Table \ref{table1}.

{\tiny
\begin{table}[ht]\centering
\caption{Comparison of total number of CMOS transistors (4-bit).}\label{table1}
\resizebox{\textwidth}{!}{\begin{tabular}{|l|c|c|c|}
\hline
\textbf{\begin{tabular}[c]{@{}l@{}}Name of Adder\\ (transistors / block)\end{tabular}} & \textbf{\begin{tabular}[c]{@{}c@{}}Carry-select\\ adder\end{tabular}}  & \textbf{\begin{tabular}[c]{@{}c@{}}Carry-select adder\\ with binary to\\ excess 1 converter\end{tabular}} & \textbf{\begin{tabular}[c]{@{}c@{}}Half adder based carry-\\ select adder with binary\\ to excess-1 converter\end{tabular}} \\ \hline
Total Half Adders (12/HA)                                                              & 0                                                                      & 0                                                                                                         & 4                                                                                                                           \\ \hline
Total Full Adders (30/FA )                                                             & 8                                                                      & 4                                                                                                         & 0                                                                                                                           \\ \hline
Total MUXs (2/MUX)                                                                     & 5                                                                      & 5                                                                                                         & 8                                                                                                                           \\ \hline
Total BEC-1s (30/4-bit)                                                                & 0                                                                      & 1                                                                                                         & 0                                                                                                                           \\ \hline
Total BEC-1s (8/2-bit)                                                                 & 0                                                                      & 0                                                                                                         & 8                                                                                                                           \\ \hline
\textbf{\begin{tabular}[c]{@{}l@{}}Total   Number of\\ Transistors\end{tabular}}       & \textbf{\begin{tabular}[c]{@{}c@{}}(8*30)+(5*2)\\  = 250\end{tabular}} & \textbf{\begin{tabular}[c]{@{}c@{}}(4*30)+(5*2)+(8*2)\\ = 142\end{tabular}}                               & \textbf{\begin{tabular}[c]{@{}c@{}}(4*12)+(8*2)+(8*8)\\ = 128\end{tabular}}                                                 \\ \hline
\end{tabular}}
\end{table}
}

Furthermore, in Figure \ref{fig5a}, the gate-level circuit diagram of 1-bit of our proposed adder is shown, and the working of the same circuit is explained in Figure \ref{fig5b}. This 1-bit version (used in the proposed multiplier design) of our proposed adder works as a FA but with high speed and better energy efficiency. In this circuit (Figure \ref{fig5a}), the incoming carry bit contributes to the final sum and carry-out with only one transistor delay. In some cases, i.e., in the Dadda circuit, the incoming carry bit may be late as compared to inputs A and B. For example, in Figure \ref{fig6}, the circuit block diagram of the 4-bit proposed multiplier is shown, where an HA-CSA-BEC-1 block is outlined in red. The inputs A and B of that block come directly from AND gates, i.e., the PPs generator; however, the carry-in bit comes through another adder block, which adds some delay to this bit as compared to the A and B inputs. This gap of time adds some delay to the final results. However, in the proposed adder (Figure \ref{fig5a}), the incoming carry bit can cover this delay, and the next adder doesn’t have to wait for carry-in until Step-3 (Figure \ref{fig5b}), because the carry-in bit contributes only at the last stage, whereas A and B have to pass through Step-1 and/or Step-2, depending on the actual value of Cin. The output expression of the proposed adder is given by
\begin{equation} \label{outputexpressions}
\begin{array}{l}
{\rm{outpu}}{{\rm{t}}_{{\rm{sum}}}} = A \oplus B\,\,{\rm{and}}\,\,{\rm{outpu}}{{\rm{t}}_{{\rm{carry}}}} = A \circ B\,\,\,{\rm{if}}\,\,\,{C_{in}} = {\rm{ }}0,\\
{\rm{outpu}}{{\rm{t}}_{{\rm{sum}}}} = \overline {A \oplus B} \,\,{\rm{and}}\,\,{\rm{outpu}}{{\rm{t}}_{{\rm{carry}}}} = \left( {A \circ B} \right) \oplus \left( {A \oplus B} \right)\,\,\,{\rm{if}}\,\,\,{C_{in}} = {\rm{ }}1.
\end{array}
\end{equation}

\begin{subfigures}

\begin{figure}[h!]
\centerline{
\includegraphics[width=0.5\textwidth]{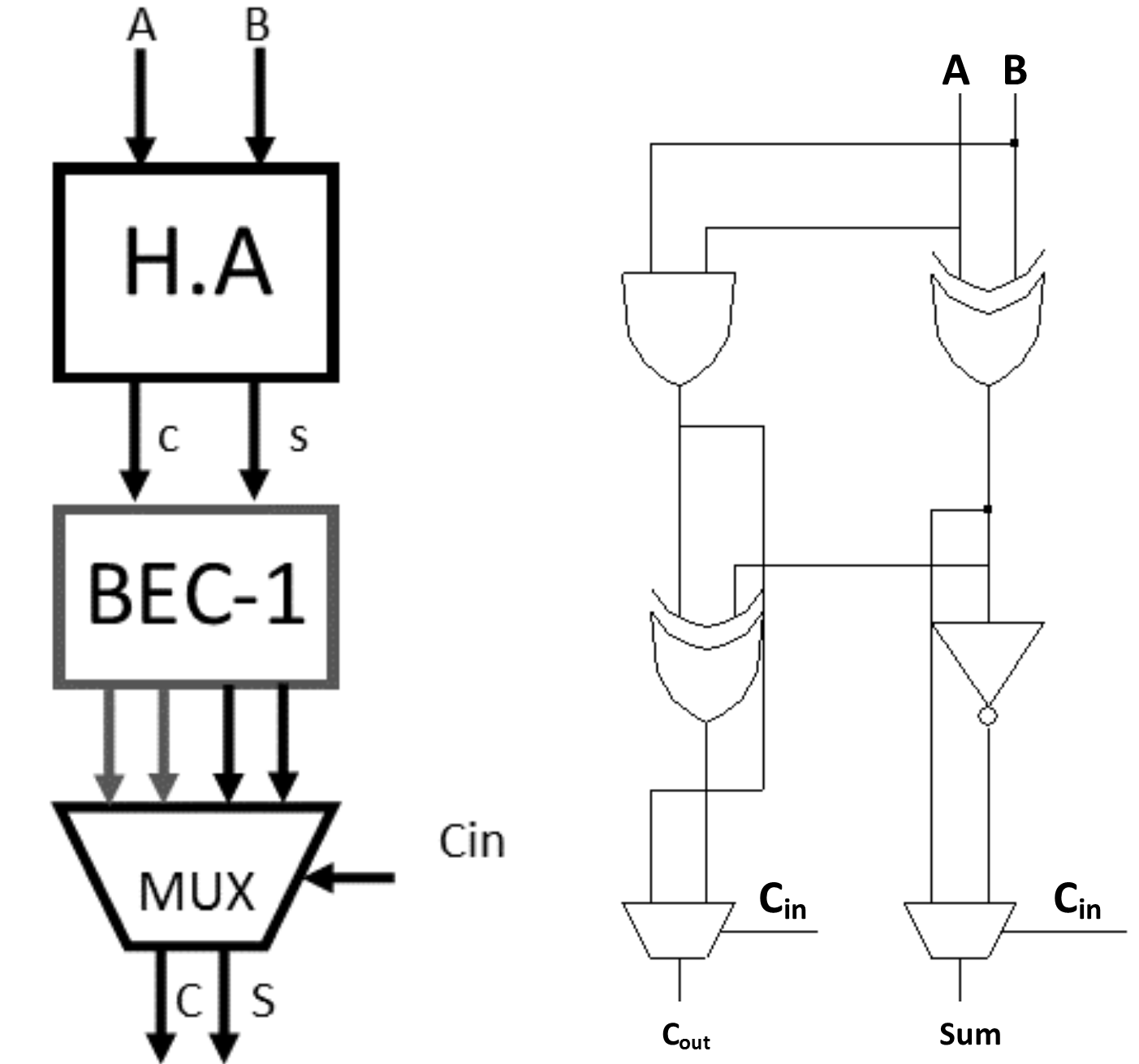}}
\caption{1-bit HA based CSA with BEC1 (\textbf{Left}: block level circuit diagram, \textbf{Right}: Gate level circuit diagram).}
\label{fig5a}
\end{figure}

\begin{figure}[h!]
\centerline{
\includegraphics[width=1\textwidth]{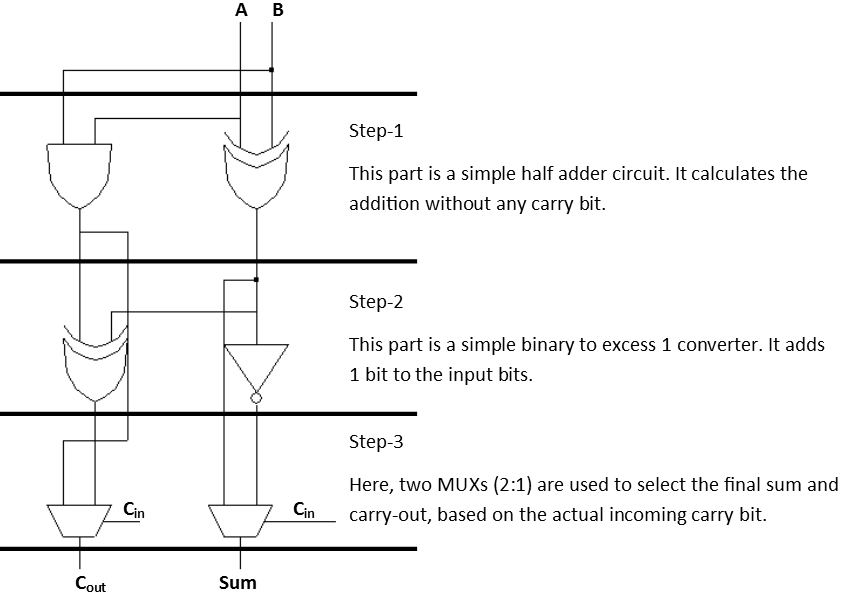}}
\caption{HA based CSA with BEC1, explained.}
\label{fig5b}
\end{figure}

\end{subfigures}

The 1-bit circuit of the proposed adder is simulated using a TSMC 50-nm library. The layout is tested on different frequencies, i.e., 0.5 GHz, 1 GHz, 2 GHz, and 3.33 GHz, and the results, i.e., delay, power consumption, and power-delay product (PDP), are shown in Table \ref{table2}.

\begin{table}[h!]\centering
\caption{Proposed adder simulation results with TSMC 50 nm.}\label{table2}
\resizebox{\textwidth}{!}{
\begin{tabular}{|lcccc|}
\hline
\multicolumn{5}{|c|}{\textbf{HA Based Carry-select Adder with Binary to Excess-1 Converter ( TSMC 50 nm )}}                                                                              \\ \hline
\multicolumn{1}{|l|}{\textbf{Frequency}}         & \multicolumn{1}{c|}{\textbf{500 MHz}} & \multicolumn{1}{c|}{\textbf{1 GHz}} & \multicolumn{1}{c|}{\textbf{2 GHz}} & \textbf{3.33 GHz} \\ \hline
\multicolumn{1}{|l|}{\textbf{Delay}}             & \multicolumn{1}{c|}{14 ps}             & \multicolumn{1}{c|}{14 ps}           & \multicolumn{1}{c|}{14 ps}           & 14 ps              \\ \hline
\multicolumn{1}{|l|}{\textbf{Power Consumption}} & \multicolumn{1}{c|}{0.779 uW}          & \multicolumn{1}{c|}{1.6 uW}          & \multicolumn{1}{c|}{3.28 uW}         & 5.7 uW             \\ \hline
\multicolumn{1}{|l|}{\textbf{Energy (PDP)}}      & \multicolumn{1}{c|}{0.0109 fJ}         & \multicolumn{1}{c|}{0.0224 fJ}       & \multicolumn{1}{c|}{0.0459 fJ}       & 0.0798 fJ          \\ \hline
\multicolumn{1}{|l|}{\textbf{Transistors count}} & \multicolumn{1}{c|}{24}               & \multicolumn{1}{c|}{24}             & \multicolumn{1}{c|}{24}             & 24                \\ \hline
\end{tabular}}
\end{table}

\begin{figure}[h!]
\centerline{
\includegraphics[width=1\textwidth]{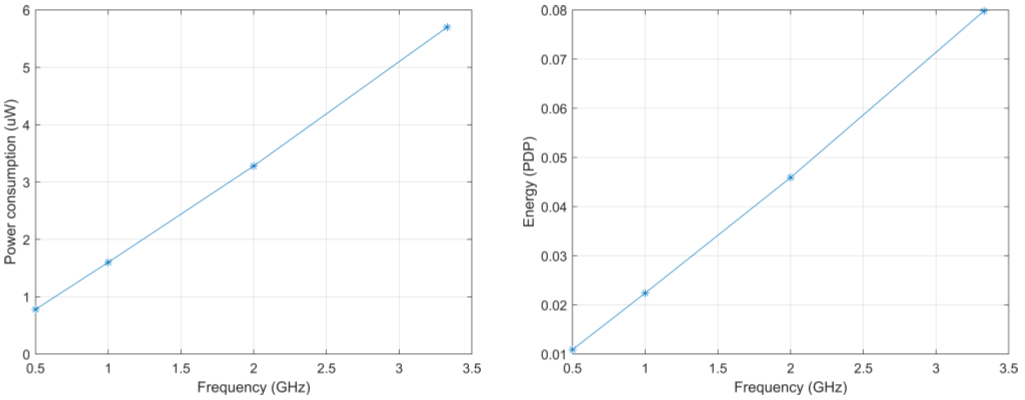}}
\caption{Energy (PDP) and power consumption of the proposed adder versus the frequency with TSMC 50 nm.}
\label{fig5b}
\end{figure}

A 1-bit version of our proposed HA-CSA-BEC1 consumes 0.779, 1.6, 3.28, and 5.7 uW  of power at 0.5, 1, 2, and 3.33 GHz,, respectively. Furthermore, it can be seen from Table \ref{table2} that the delay is 14 ps and the total number of transistors needed for the implementation is 24. Similarly, the PDP is 0.0109, 0.0224, 0.0459, and 0.0798 (fJ) at 0.5, 1, 2, and 3.33 GHz, respectively.

\subsection{Proposed full Dadda multiplier with HA-CSA-BEC1}
\begin{figure}[h!]
\centerline{
\includegraphics[width=0.7\textwidth]{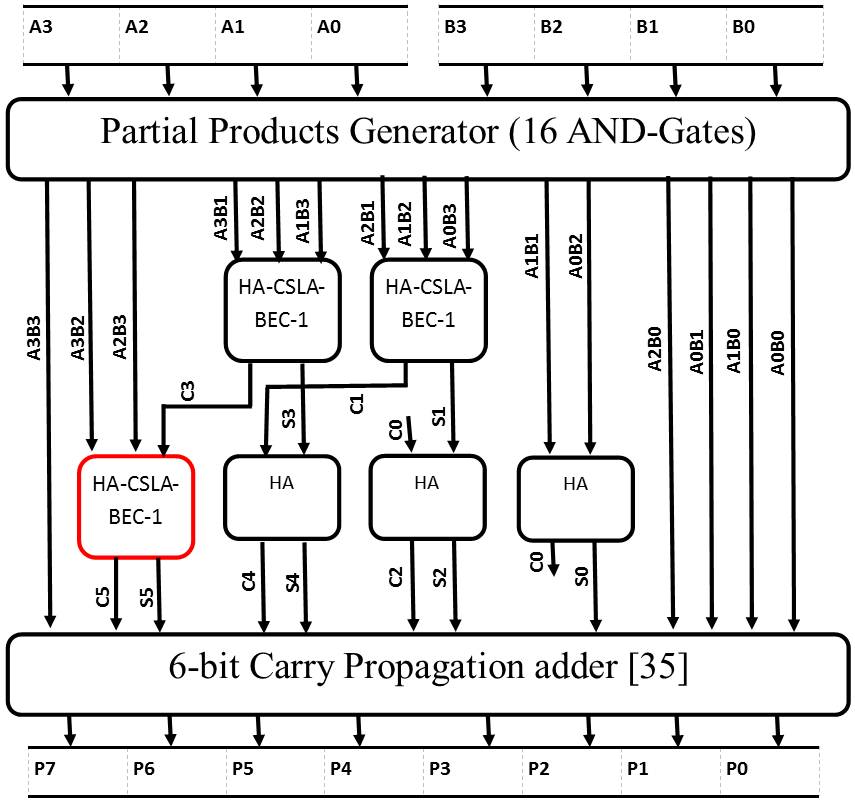}}
\caption{The circuit-block diagram of proposed multiplier Circuit (4-bit).}
\label{fig6}
\end{figure}

The circuit-block diagram of a proposed 4-bit multiplier is shown in Figure \ref{fig6}. The CMOS layouts of 4-bit and 8-bit are depicted in Figures \ref{fig7a}, and \ref{fig7b}, respectively.

In Figure \ref{fig6}, there are two 4-bit numbers available for the input of the PPs-Generator, which results in 16 multiplications, i.e., 16 AND gates. These multiplications are added to each other by following the full Dadda algorithm, i.e., Figure \ref{fig2a}. By following the layer-reduction pattern of full Dadda, two 6-bit layers are obtained, which are then fed into a carry propagation adder \cite{ref35} to compute the final 8-bit product. Figure 6 provides two 4-bit numbers as inputs to the PPs-Generator, resulting in 16 multiplications or AND gates. These multiplications are then combined using the full Dadda algorithm (Figure 2a) to obtain two 6-bit layers. Finally, a carry propagation adder [35] is utilized to calculate the 8-bit product.

\begin{subfigures}

\begin{figure}[h!]
\centerline{
\includegraphics[width=0.58\textwidth]{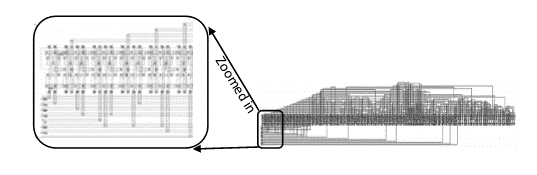}}
\caption{Proposed multiplier layout using TSMC 50 nm (4-bit).}
\label{fig7a}
\end{figure}

\begin{figure}[h!]
\centerline{
\includegraphics[width=0.58\textwidth]{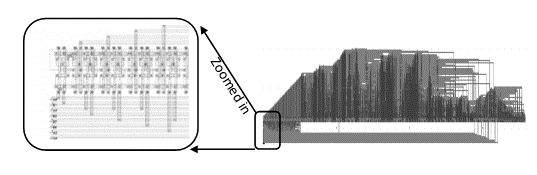}}
\caption{Proposed multiplier layout using TSMC 50 nm (8-bit).}
\label{fig7b}
\end{figure}

\end{subfigures}

For an \(M\)-bit proposed multiplier, the total number of HA-CSA-BEC1s, HAs, AND-Gates (PPs), and CPA size can be calculated using the following equations:

\begin{equation}
\label{eq4}
{\rm{HA\rm{-}CSA\rm{-}BEC1s}} = (M - 1) \times (M - 3),{\rm{valid\,for\, }}M > 2,
\end{equation}

\begin{equation}
\label{eq5}
{\rm{HAs}} = M - 1,
\end{equation}

\begin{equation}
\label{eq6}
{\rm{AND\,gates\,for\,PPs}} = {M^2},
\end{equation}

\begin{equation}
\label{eq7}
{\rm{Size\,of\,CPA}} = 2 \times (M - 2).
\end{equation}

To calculate the total number of MUXs (2:1) in an M-bit proposed multiplier, following equation can be used:

\begin{equation}
\label{eq8}
{\rm{MUXs}} = 2 \times (M - 1) \times (M - 3).
\end{equation}

\section{Simulation results and performance comparison}
As aforementioned, the proposed multiplier layout is designed and simulated in Microwind software using TSMC 50 nm, 90 nm, and 120 nm CMOS technologies on different frequencies, i.e., 0.5 GHz, 1 GHz, 2 GHz, and 3.33 GHz, and results are tabulated in the Tables \ref{table3}, \ref{table4}, and \ref{table5}.

\begin{table}[h!]\centering\caption{Simulation results of proposed multiplier with proposed adder (TSMC 50 nm).}\label{table3}
\resizebox{\textwidth}{!}{
\begin{tabular}{|l|cccc|}
\hline
\textbf{\textcolor{black}{TSMC   50 nm}}           & \multicolumn{4}{c|}{\textbf{Proposed Full Dadda with proposed adder}}                                                 \\ \hline
\textbf{Frequency (GHz)}        & \multicolumn{1}{c|}{\textbf{0.5}} & \multicolumn{1}{c|}{\textbf{1}} & \multicolumn{1}{c|}{\textbf{2}} & \textbf{3.33} \\ \hline
\textbf{Delay (ps)}             & \multicolumn{1}{c|}{75}           & \multicolumn{1}{c|}{75}         & \multicolumn{1}{c|}{76}         & 76            \\ \hline
\textbf{Power Consumption (uW)} & \multicolumn{1}{c|}{5.341}        & \multicolumn{1}{c|}{9.083}      & \multicolumn{1}{c|}{15.9}       & 25            \\ \hline
\textbf{Energy (fJ)}            & \multicolumn{1}{c|}{0.40057}      & \multicolumn{1}{c|}{0.68122}    & \multicolumn{1}{c|}{1.2084}     & 1.9           \\ \hline
\textbf{Transistors Count}      & \multicolumn{1}{c|}{338}          & \multicolumn{1}{c|}{338}        & \multicolumn{1}{c|}{338}        & 338           \\ \hline
\end{tabular}}
\end{table}

\begin{table}[h!]\centering\caption{Simulation results of proposed multiplier with proposed adder (TSMC 90 nm).}\label{table4}
\resizebox{\textwidth}{!}{\begin{tabular}{|l|cccc|}
\hline
\textbf{\textcolor{black}{TSMC   90 nm}}           & \multicolumn{4}{c|}{\textbf{Proposed Full Dadda with proposed   adder}}                                               \\ \hline
\textbf{Frequency (GHz)}        & \multicolumn{1}{c|}{\textbf{0.5}} & \multicolumn{1}{c|}{\textbf{1}} & \multicolumn{1}{c|}{\textbf{2}} & \textbf{3.33} \\ \hline
\textbf{Delay (ps)}             & \multicolumn{1}{c|}{99}           & \multicolumn{1}{c|}{99}         & \multicolumn{1}{c|}{99}         & 100           \\ \hline
\textbf{Power Consumption (mW)} & \multicolumn{1}{c|}{0.0728}       & \multicolumn{1}{c|}{0.121}      & \multicolumn{1}{c|}{0.213}      & 0.316         \\ \hline
\textbf{Energy (pJ)}            & \multicolumn{1}{c|}{7.21}         & \multicolumn{1}{c|}{11.979}     & \multicolumn{1}{c|}{21.087}     & 31.6          \\ \hline
\textbf{Transistors Count}      & \multicolumn{1}{c|}{338}          & \multicolumn{1}{c|}{338}        & \multicolumn{1}{c|}{338}        & 338           \\ \hline
\end{tabular}}
\end{table}

\begin{table}[h!]\centering\caption{Simulation results of proposed multiplier with proposed adder (TSMC 120 nm).}\label{table5}
\resizebox{\textwidth}{!}{\begin{tabular}{|l|cccc|}
\hline
\textbf{\textcolor{black}{TSMC   120 nm}}           & \multicolumn{4}{c|}{\textbf{Proposed Full Dadda with proposed   adder}}                                               \\ \hline
\textbf{Frequency (GHz)}        & \multicolumn{1}{c|}{\textbf{0.5}} & \multicolumn{1}{c|}{\textbf{1}} & \multicolumn{1}{c|}{\textbf{2}} & \textbf{3.33} \\ \hline
\textbf{Delay (ps)}             & \multicolumn{1}{c|}{123}          & \multicolumn{1}{c|}{123}        & \multicolumn{1}{c|}{123}        & 123           \\ \hline
\textbf{Power Consumption (mW)} & \multicolumn{1}{c|}{0.119}        & \multicolumn{1}{c|}{0.185}      & \multicolumn{1}{c|}{0.303}      & 0.483         \\ \hline
\textbf{Energy (pJ)}            & \multicolumn{1}{c|}{14.637}       & \multicolumn{1}{c|}{22.755}     & \multicolumn{1}{c|}{37.269}     & 59.409        \\ \hline
\textbf{Transistors Count}      & \multicolumn{1}{c|}{338}          & \multicolumn{1}{c|}{338}        & \multicolumn{1}{c|}{338}        & 338           \\ \hline
\end{tabular}}
\end{table}

According to the simulation results in Table \ref{table3}, the proposed multiplier with TSMC 50 nm, consumes 5.34, 9.083, 15.9, and 25 uW of power at 0.5, 1, 2, and 3.33 GHz frequencies, respectively. The delay is 75 ps, and the energy usage results are 0.4, 0.681, 1.208, and 1.9 fJ at 0.5, 1, 2, and 3.33 GHz, respectively. Similarly, the simulation results of the proposed design with TSMC 90 nm and TSMC 120 nm can be seen in Tables \ref{table4} and \ref{table5}, respectively.

The proposed design is also compared to some recent, closely related works in the literature, i.e., \cite{ref16,ref17,ref43,ref44,ref46,ref47,ref48,ref49}, where \cite{ref49} is an approximate multiplier-based design. It can be seen in Table \ref{table6} that our design outperforms these in terms of delay, power consumption, energy efficiency, and transistor count. It is important to notice that each paper is being compared separately with the proposed one. The reason is that each research paper’s results have been simulated under different frequencies and technologies, i.e., 45 nm, 90 nm, and 120 nm, etc. Therefore, to make a justified comparison, our design is simulated with the same frequencies and technologies. For example, in \cite{ref17}, TSMC-90 nm technology with a 500 MHz frequency is used, while in \cite{ref46} TSMC-40 nm technology with a 100 MHz frequency is used, and to make a comparison with these papers, different technologies and frequencies are used correspondingly. 

\begin{table}[h!]\centering\caption{Proposed multiplier comparison with recent works (4-bit).}\label{table6}
\resizebox{\textwidth}{!}{
\color{black}\begin{tabular}{|l|l|l|l|l|l|l|l|l|}
\hline
\textbf{SI. No.} & \multicolumn{1}{c|}{\textbf{Reference}}                             & \multicolumn{1}{c|}{\textbf{\begin{tabular}[c]{@{}c@{}}CMOS Design\\ Technology\end{tabular}}} & \multicolumn{1}{c|}{\textbf{Delay}}                    & \multicolumn{1}{c|}{\textbf{\begin{tabular}[c]{@{}c@{}}Power\\ @Frequency\end{tabular}}} & \multicolumn{1}{c|}{\textbf{\begin{tabular}[c]{@{}c@{}}PDP\\ (Energy)\end{tabular}}} & \multicolumn{1}{c|}{\textbf{\begin{tabular}[c]{@{}c@{}}Transistors/\\ Area\end{tabular}}}                & \multicolumn{1}{c|}{\textbf{\begin{tabular}[c]{@{}c@{}}Energy-delay\\  product\\ (EDP)\end{tabular}}} & \multicolumn{1}{c|}{\textbf{\begin{tabular}[c]{@{}c@{}}Normalized\\ EDP\end{tabular}}} \\ \hline
\textbf{1.}      & \begin{tabular}[c]{@{}l@{}}{\cite{ref17}}  \\ {[}Proposed{]}\end{tabular} & \begin{tabular}[c]{@{}l@{}}TSMC 90nm \\ TSMC 90nm\end{tabular}                                 & \begin{tabular}[c]{@{}l@{}}200 ps\\ 99 ps\end{tabular} & \begin{tabular}[c]{@{}l@{}}0.109 mW@500MHz\\ 0.0728 mW@500MHz\end{tabular}               & \begin{tabular}[c]{@{}l@{}}21.80 fJ\\ 7.21 fJ\end{tabular}                           & \begin{tabular}[c]{@{}l@{}}472 / NA\\ 338 / 3220 \(\mu {{\rm{m}}^2}\)\end{tabular}                       & \begin{tabular}[c]{@{}l@{}}4.3600e-24\\ 7.1379e-25\end{tabular}                                       & \begin{tabular}[c]{@{}l@{}}0.10\\ 0.01\end{tabular}                                    \\ \hline
\textbf{2.}      & \begin{tabular}[c]{@{}l@{}}{\cite{ref16}} \\ {[}Proposed{]}\end{tabular}  & \begin{tabular}[c]{@{}l@{}}TSMC 90nm\\ TSMC 90nm\end{tabular}                                  & \begin{tabular}[c]{@{}l@{}}262 ps\\ 99 ps\end{tabular} & \begin{tabular}[c]{@{}l@{}}0.623 mW@3.5GHz\\ 0.342 mW@3.5GHz\end{tabular}                & \begin{tabular}[c]{@{}l@{}}163.22 fJ\\ 33.858 fJ\end{tabular}                        & \begin{tabular}[c]{@{}l@{}}NA / 7752 \(\mu {{\rm{m}}^2}\)\\ 338 / 3220 \(\mu {{\rm{m}}^2}\)\end{tabular} & \begin{tabular}[c]{@{}l@{}}4.2764e-23\\ 3.3519e-24\end{tabular}                                       & \begin{tabular}[c]{@{}l@{}}1.00\\ 0.07\end{tabular}                                    \\ \hline
\textbf{3.}      & \begin{tabular}[c]{@{}l@{}}{[}43{]}\\ {[}Proposed{]}\end{tabular}   & \begin{tabular}[c]{@{}l@{}}TSMC 90nm\\ TSMC 90nm\end{tabular}                                  & \begin{tabular}[c]{@{}l@{}}376 ps\\ 99 ps\end{tabular} & \begin{tabular}[c]{@{}l@{}}0.2015 mW@500MHz\\ 0.0728 mW@500MHz\end{tabular}              & \begin{tabular}[c]{@{}l@{}}75.760 fJ\\ 7.210 fJ\end{tabular}                         & \begin{tabular}[c]{@{}l@{}}NA / 3100 \(\mu {{\rm{m}}^2}\)\\ 338 / 3220 \(\mu {{\rm{m}}^2}\)\end{tabular} & \begin{tabular}[c]{@{}l@{}}2.8486e-23\\ 7.1379e-25\end{tabular}                                       & \begin{tabular}[c]{@{}l@{}}0.16\\ 0.01\end{tabular}                                    \\ \hline
\textbf{4.}      & \begin{tabular}[c]{@{}l@{}}{\cite{ref43}}\\ {[}Proposed{]}\end{tabular}   & \begin{tabular}[c]{@{}l@{}}TSMC 32nm\\ TSMC 50nm\end{tabular}                                  & \begin{tabular}[c]{@{}l@{}}295 ps\\ 75 ps\end{tabular} & \begin{tabular}[c]{@{}l@{}}3.60 uW@N/A\\ 2.42 uW@125MHz\end{tabular}                     & \begin{tabular}[c]{@{}l@{}}1.060 fJ\\ 0.181 fJ\end{tabular}                          & \begin{tabular}[c]{@{}l@{}}NA / NA\\ 338 / 1540 \(\mu {{\rm{m}}^2}\)\end{tabular}                        & \begin{tabular}[c]{@{}l@{}}3.1270e-25\\ 1.3575e-26\end{tabular}                                       & \begin{tabular}[c]{@{}l@{}}0.007\\ 0.003\end{tabular}                                  \\ \hline
\textbf{5.}      & \begin{tabular}[c]{@{}l@{}}{\cite{ref46}} \\ {[}Proposed{]}\end{tabular}  & \begin{tabular}[c]{@{}l@{}}TSMC 40nm\\ TSMC 50nm\end{tabular}                                  & \begin{tabular}[c]{@{}l@{}}112 ps\\ 75 ps\end{tabular} & \begin{tabular}[c]{@{}l@{}}3.25 uW@100MHz\\ 2.42 uW@125MHz\end{tabular}                  & \begin{tabular}[c]{@{}l@{}}0.364 fJ\\ 0.181 fJ\end{tabular}                          & \begin{tabular}[c]{@{}l@{}}532 / NA\\ 338 / 1540 \(\mu {{\rm{m}}^2}\)\end{tabular}                       & \begin{tabular}[c]{@{}l@{}}4.0768e-26\\ 1.3575e-26\end{tabular}                                       & \begin{tabular}[c]{@{}l@{}}0.0008\\ 0.0003\end{tabular}                                \\ \hline
\textbf{6.}      & \begin{tabular}[c]{@{}l@{}}{\cite{ref47}} \\ {[}Proposed{]}\end{tabular}  & \begin{tabular}[c]{@{}l@{}}TSMC 40nm\\ TSMC 50nm\end{tabular}                                  & \begin{tabular}[c]{@{}l@{}}119 ps\\ 75 ps\end{tabular} & \begin{tabular}[c]{@{}l@{}}4.008 uW@100MHz\\ 2.42 uW@125MHz\end{tabular}                 & \begin{tabular}[c]{@{}l@{}}0.476 fJ\\ 0.181 fJ\end{tabular}                          & \begin{tabular}[c]{@{}l@{}}579 / NA\\ 338 / 1540 \(\mu {{\rm{m}}^2}\)\end{tabular}                       & \begin{tabular}[c]{@{}l@{}}5.6644e-26\\ 1.3575e-26\end{tabular}                                       & \begin{tabular}[c]{@{}l@{}}0.009\\ 0.0003\end{tabular}                                 \\ \hline
\textbf{7.}      & \begin{tabular}[c]{@{}l@{}}{\cite{ref48}}  \\ {[}Proposed{]}\end{tabular} & \begin{tabular}[c]{@{}l@{}}TSMC 40nm\\ TSMC 50nm\end{tabular}                                  & \begin{tabular}[c]{@{}l@{}}125 ps\\ 75 ps\end{tabular} & \begin{tabular}[c]{@{}l@{}}5.26 uW@100MHz\\ 2.42 uW@125MHz\end{tabular}                  & \begin{tabular}[c]{@{}l@{}}0.657 fJ\\ 0.181 fJ\end{tabular}                          & \begin{tabular}[c]{@{}l@{}}539 / NA\\ 338 / 1540 \(\mu {{\rm{m}}^2}\)\end{tabular}                       & \begin{tabular}[c]{@{}l@{}}8.2125e-26\\ 1.3575e-26\end{tabular}                                       & \begin{tabular}[c]{@{}l@{}}0.0011\\ 0.0003\end{tabular}                                \\ \hline
\textbf{8.}      & \begin{tabular}[c]{@{}l@{}}{\cite{ref49}}\\ {[}Proposed{]}\end{tabular}   & \begin{tabular}[c]{@{}l@{}}TSMC 45nm\\ TSMC 50nm\end{tabular}                                  & \begin{tabular}[c]{@{}l@{}}49 ps\\ 75 ps\end{tabular}  & \begin{tabular}[c]{@{}l@{}}26 uW @ N/A\\ 2.42 uW @123MHz\end{tabular}                    & \begin{tabular}[c]{@{}l@{}}1.3 fJ\\ 0.181 fJ\end{tabular}                            & \begin{tabular}[c]{@{}l@{}}NA / NA\\ 338 / 1540 \(\mu {{\rm{m}}^2}\)\end{tabular}                        & \begin{tabular}[c]{@{}l@{}}6.3700e-26\\ 1.3575e-26\end{tabular}                                       & \begin{tabular}[c]{@{}l@{}}0.0008\\ 0.0003\end{tabular}                                \\ \hline
\end{tabular}}
\end{table}

\section{Conclusion}
In this paper, a low-power, high-speed, and area-energy-efficient digital multiplier is designed that is based on the full Dadda algorithm and uses a new proposed full adder named a half-adder-based carry-select adder with a binary to excess-1 converter. Specifically, the circuit, CSA-BEC1 [17], is modified by replacing the FAs with HAs and transferring the overhead of carry propagation to the multiplexer stage. The performance of the proposed design is highly dependent on the HA-CSA-BEC1, which uses only 24 transistors to work as a CMOS full adder. The proposed multiplier’s circuit and layout are designed and simulated in DSCH and Microwind software, respectively, using TSMC 50 nm, 90 nm, and 120 nm with GHz frequencies, i.e., 0.5, 1, 2, and 3.33. According to the simulation results, a 4-bit proposed multiplier in TSMC-50 nm consumes 5.341 uW, 9.08 uW, 15.9 uW, and 25 uW of power at 0.5 GHz, 1 GHz, 2 GHz, and 3 GHz of frequencies, respectively, with a delay of 75 ns and an area of 338 transistors. Compared to many recent related works, the proposed design uses a smaller number of transistors and has a lower delay, PDP, and power consumption. The proposed design can be a good candidate for resource-limited digital control applications.


\end{document}